\documentclass[useAMS,usenatbib,usegraphicx]{mn2e}
\usepackage{epsf, epsfig}
\usepackage{Times}

\def\be{\begin{equation}}
\def\ee{\end{equation}}

\def\be{\begin{equation}}
\def\ee{\end{equation}}
\catcode`\@=11 % This allows us to modify PLAIN macros.
\def\@versim#1#2{\vcenter{\offinterlineskip
\ialign{$\m@th#1\hfil##\hfil$\crcr#2\crcr\sim\crcr } }}
\def\lsim{\mathrel{\mathpalette\@versim<}}
\def\gsim{\mathrel{\mathpalette\@versim>}}

\newcommand{\arcdeg}{\ensuremath{^{\circ}}}

\title[Modeling the Broadband SED of XTE~J1550--564 and H~1743--322]
{Modeling the Broadband Spectral Energy Distribution of the Microquasars XTE~J1550--564 and H~1743--322}
\author[Xue, Wu, \& Cui]
{Yongquan Xue$^{1}$\thanks{E-mail: xuey@physics.purdue.edu}, 
Xue-Bing Wu$^{2}$\thanks{E-mail: wuxb@vega.bac.pku.edu.cn}, 
Wei Cui$^{2}$\thanks{On sabbatical leave from the Department of Physics, 
Purdue University; e-mail: cui@physics.purdue.edu} 
\\
$^{1}$\/Department of Physics, Purdue University, West Lafayette, IN 47907\\
$^{2}$\/Department of Astronomy, Peking University, Beijing 100871, P. R. China}

\begin{document}

\date{Received 2007}

\pagerange{\pageref{firstpage}--\pageref{lastpage}} \pubyear{2007}

\maketitle

\label{firstpage}

\begin{abstract}
We report results from a systematic study of the spectral energy distribution
(SED) and spectral evolution of XTE~J1550--564 and H~1743--322 in outburst. 
The jets of both sources have been 
directly imaged at both radio and X-ray frequencies, which makes it possible 
to constrain the spectrum of the radiating electrons in the jets. We modelled 
the observed SEDs of the jet `blobs' with synchrotron emission alone and 
with synchrotron emission plus inverse Compton scattering. The results favor 
a pure synchrotron origin of the observed jet emission. Moreover, we found
evidence that the shape of the electron spectral distribution is similar for 
all jet `blobs' seen. Assuming that this is the case for the jet as a whole,
we then applied the synchrotron model to the radio spectrum of the total 
emission and extrapolated the results to higher frequencies. In spite of 
significant degeneracy in the fits, it seems clear that, while the synchrotron 
radiation from the jets can account for nearly 100\% of the measured radio 
fluxes, it contributes little to the observed X-ray emission, when the source 
is relatively bright. In this case, the X-ray emission is most likely 
dominated by emission from the accretion flows. When the source becomes 
fainter, however, the jet emission becomes more important, even dominant, 
at X-ray energies. We also examined the spectral properties of the sources
during outbursts and the correlation between the observed radio and 
X-ray variabilities. The implication of the results is discussed.
\end{abstract}

\begin{keywords}
X-rays: binaries --- X-rays: individual (XTE J1550--564, H 1743--322) --- 
black hole physics  --- accretion, accretion discs
\end{keywords}

\section{Introduction}

There is growing evidence that the central engine of microquasars is
qualitatively similar to that of active galactic nuclei (AGN) or of gamma-ray
bursts (GRBs). Collectively, these systems provide an excellent laboratory
for studying particle acceleration in the jets of black holes over a vast
range of physical scales (Mirabel 2004; Cui 2006). As has been well 
demonstrated in the
study of AGN and GRBs, modelling the broadband spectral energy distribution
(SED) of microquasars can be a very effective way to cast light on emission 
mechanisms and thus on the nature of the central engine. Unlike the cases of 
AGN and GRBs, however, the availability of such data is highly limited for 
microquasars, due to the combination of the transient nature of most such 
sources and the rarity of their outbursting activities (during which they 
can usually be seen). In spite of the difficulty, much has been learned from 
the studies that have been carried out (e.g., Markoff,
Falcke \& Fender 2001; Hynes et al. 2002; Ueda 
et al. 2002; Chaty et al. 2003; Fuchs et al. 2003; Markoff \& Nowak 2004;
Markoff, Nowak \& Wilms 2005; Yuan, Cui \& Narayan 2005).

The broadband SED of microquasars is usually composed of several distinct
components. Very roughly, at radio frequencies, the spectrum can be described
by a power law, which may have either positive or negative frequency
dependence. The emission is generally thought to be synchrotron radiation
from relativistic electrons in the jets, which may be optically thin or thick
to synchrotron self-absorption (Hjellming \& Han 1995; Fender 2006). Such a
synchrotron spectrum could extend up to infrared/optical frequencies in some
cases (e.g., Chaty et al. 2003; Migliari et al. 2007). From optical to soft X-ray frequencies, the
spectrum takes on a blackbody-like shape, which is usually viewed as the
signature of an optically thick, geometrically thin accretion disc (review
by Liang 1998). It is seen to peak at different frequencies for different
sources or even for a given source at different fluxes; the latter has been
interpreted as being related to the movement of the inner edge of the disc
(e.g., Esin, McClintock \& Narayan 1997). 
From hard X-ray to soft gamma-ray frequencies, the
spectrum can, once again, be roughly described by a power law, which rolls 
over at some characteristic frequency under certain circumstances. It is 
generally modelled as
Compton upscattering of soft photons by energetic electrons in an
optically-thin configuration (Liang 1998).

The physical origin of Comptonizing electrons is still not entirely clear.
They may be thermal electrons associated with advection dominated accretion
flows (ADAF; see, e.g., Narayan, Mahadevan \& Quataert 1998) or magnetic flares (e.g.,
Poutanen \& Fabian 1999) above the thin disc. Over the years, intense efforts
have been made to fit the X-ray spectrum of microquasars in the so-called
low-hard state with models that include optically-thick emission from the
thin disc and a Comptonized emission from some `corona'.
The most successful and physically self-consistent models all seem to prefer
a geometry that consists of an inner (roughly) spherical corona plus an outer
thin disc (e.g., Dove, Wilms \& Begelman 1997; Esin et al. 1997). Alternatively, the
Comptonizing electrons may be non-thermal in nature (e.g., Zdziarski et al. 
2001; Titarchuk \& Shrader 2002). This is thought to be
particularly relevant to the so-called high-soft state of microquasars
(Grove et al. 1998). A promising place to accelerate electrons to the
required energies is the jets, as in the case of AGN and GRBs, although the
electrons could also be associated with the accretion flows. The broadband
SED has proven to be a valuable tool for assessing the roles of jets and
accretion flows in microquasars, as well as the coupling between the two
(Yuan et al. 2005).

In this work, we examined two microquasars, XTE J1550--564 and H 1743--322,
whose jets have been directly imaged at both radio and X-ray frequencies 
(Corbel et al. 2002, 2005), mainly to gain insights into the roles of the 
jets and accretion flows in microquasars. The broadband SED of the jets 
allowed us to assess the contribution of jets to 
the overall SED in a relatively model-independent manner. Moreover, modelling 
the jet SED helped constrain the properties of radiating electrons in the
jets and thus reduced degeneracy in the modelling of the overall SED of the 
source. We also studied the spectral evolution of the sources during 
outburst and the correlation between the observed radio and X-ray 
variabilities.

\section{X-ray Data}

\subsection{{\em Chandra} Data}

We used the {\em Chandra} data to constrain the X-ray emission of the jets.
We searched the {\em Chandra} archival database and found a total of 14 
observations of XTE~J1550--564 and 7 observations of H~1743--322, respectively,
all taken with the ACIS detector (and some also with the HETG inserted).
Since we are only interested in imaging observations 
here, we eliminated the ones taken in the continuous clocking mode. The 
remaining observations are shown in Table~1, along with references to the 
published works based on these observations.

\begin{table}
\caption{Log of the {\em Chandra} observations}
\begin{tabular}{ccccc} \hline \hline
 & MJD & UT Date & Jet(s)$^a$ & \\
ObsID & (Start time) & (Start time) & Detected & Ref. \\ \hline
& & XTE J1550--564 & & \\ \hline
679 & 51704.49 & 06/09/2000 & E & 1,2 \\
1845 & 51777.36 & 08/21/2000 & E & 1,2,3 \\
1846 & 51798.20 & 09/11/2000 & E & 1,2,3 \\
3448 & 52344.62 & 03/11/2002 & E,W & 1,3,4 \\
3672 & 52444.38 & 06/19/2002 & W & 3,4 \\
3807 & 52541.83 & 09/24/2002 & W & 3 \\
4368 & 52667.19 & 01/28/2003 & W & 3 \\
5190 & 52935.30 & 10/23/2003 & W & 3 \\ \hline
& & H 1743--322 & & \\ \hline
4565 & 53047.85 & 02/12/2004 & E,W & 3,5 \\
4566 & 53088.74 & 03/24/2004 & E,W & 3,5 \\
4567 & 53091.26 & 03/27/2004 & E,W & 3,5 \\ \hline
\end{tabular}
\\References: (1) Corbel et al. (2002); (2) Tomsick et al. (2003); (3) Corbel et al. (2006); (4) Kaaret et al. (2003); (5) Corbel et al. (2005).\\
Note: $^a$ E=Eastern and W=Western.
\end{table}

For this work, we chose to re-analyse all the data in a consistent manner. We 
used the standard 
{\it CIAO} analysis package (version 3.3), along with the corresponding 
calibration database (CALDB 3.2.0), and followed the {\it CIAO} Science 
Threads\footnote{see http://asc.harvard.edu/ciao/threads/index.html} in
constructing the images and spectra. We used the {\it CIAO} script 
{\it specextract} to extract the X-ray spectra of the jet `blobs' seen, 
with appropriate regions for the source (circle) and background (annulus). 
The script produces spectra for the total and background emission, as well 
as the corresponding {\it rmf} and {\it arf} files for subsequent spectral 
modelling. We carried out the initial modelling of the spectra in 
XSPEC v12.2.0ba (Arnaud 1996). 

Due to the low count rates of the jet `blobs', the data do not offer a good
constraint on interstellar absorption. In all cases, we fixed the hydrogen 
column density at the line-of-sight values ($N_H=8.97$ and 
$7.21 \times 10^{21}$ cm$^{-2}$ for XTE~J1550--564 and H~1743--322, 
respectively; Dickey \& Lockman 1990). For both sources, the X-ray spectrum 
(in the range of 0.3--8 keV)
of the jet emission can be fitted satisfactorily with a power law,
where the $\chi^2$ statistics with standard weighting was employed
for the rebinned spectra of jet `blobs' with 
more than 100 counts and the $\chi^2$ statistics 
with Churazov weighting (Churazov et al. 1996) was employed 
for the unbinned spectra of those with less than
100 counts (note that Churazov weighting is recommanded by XSPEC for
analysing a spectrum with very low counts). The results are 
summarized in Table~2. We emphasize that our goal of spectral modelling here
is to get a rough idea about the spectral behavior of the jet `blobs'.
The statistics is poor, especially in
the case of H~1743--322, but it is interesting to note that the jet spectrum 
of H~1743--322 seems to be harder than that of XTE~J1550--564. For each source,
%more meaningful for XTE~J1550--564, 
the photon indices of the jet 
`blobs' seen are consistent with being the same (within uncertainties).
To derive the intrinsic 
SED of a jet `blob', we de-absorbed the raw count spectrum, using the 
photoelectric absorption cross sections of Morrison \& McCammon (1983), and 
unfolded it.

\begin{table}
\caption{Spectral parameters for the resolved jet `blobs'}
\begin{tabular}{ccccc} \hline \hline
& Jet & Number of & Photon & \\
ObsID & Blob & Counts$^a$ & Index & Norm$^b$ \\\hline
& & XTE J1550--564 & & \\\hline
679 & E & 12 & $1.7^{+1.2}_{-1.0}$ & $4.2^{+1.0}_{-0.7}$ \\
1845 & E & 24 & $1.6^{+0.8}_{-0.6}$ & $1.5^{+0.8}_{-0.6}$ \\
1846 & E & 28 & $1.6^{+0.9}_{-0.7}$ & $1.9^{+0.8}_{-0.7}$ \\
3448 & E & 34 & $2.1^{+0.8}_{-0.7}$ & $0.6^{+0.4}_{-0.3}$ \\
& W & 415 & $1.74\pm 0.18$ & $5.0^{+0.9}_{-0.8}$ \\
3672 & W & 251 & $1.73\pm 0.25$ & $4.3^{+1.0}_{-0.9}$ \\
3807 & W & 200 & $1.93^{+0.29}_{-0.27}$ & $2.9\pm 0.7$ \\
4368 & W & 113 & $2.01^{+0.42}_{-0.39}$ & $1.9^{+0.7}_{-0.6}$ \\
5190 & W & 134 & $1.98^{+0.36}_{-0.35}$ & $1.1^{+0.4}_{-0.3}$ \\\hline
& & H 1743--322 & & \\\hline
4565 & E & 28 & $0.7\pm 0.6$ & $13.5^{+3.7}_{-3.5}$ \\
& W &  5 & $0.5^{+0.8}_{-1.0}$ & $2.1^{+3.3}_{-2.0}$ \\
4566 & E & 15 & $0.6^{+0.5}_{-0.6}$ & $6.3^{+3.6}_{-3.1}$ \\
& W & 17 & $0.3^{+1.0}_{-0.9}$ & $3.6^{+3.3}_{-2.9}$ \\
4567 & E & 27 & $0.6^{+0.6}_{-0.8}$ & $6.2^{+3.5}_{-3.0}$ \\
& W & 17 & $0.2^{+0.5}_{-0.7}$ & $3.1^{+3.2}_{-2.9}$ \\\hline
\end{tabular}
\\$^a$ Number of counts (background-subtracted) in the 0.3--8 keV energy range.\\
$^b$ For XTE J1550--564, Norm is in the unit of
$10^{-5}$ photons keV$^{-1}$ cm$^{-2}$ s$^{-1}$ at 1 keV;
for H 1743--322, Norm is in the unit of
$10^{-7}$ photons keV$^{-1}$ cm$^{-2}$ s$^{-1}$ at 1 keV.
\end{table}

\subsection{{\em RXTE} Data}

We used the {\em RXTE} data to obtain the X-ray SED of the {\em total} 
emission. XTE~J1550--564 underwent major outbursts during 1998--2000 and 
H~1743--322 during 2003--2005. We analysed a series of archival {\em RXTE} 
observations taken of them during their respective outbursts. For 
XTE~J1550--564, we selected more observations during the times of significant 
spectral variability (e.g., the rising phase of the 1998/1999 outburst; Cui 
et al. 1999) and less when the spectral variability of the source is less 
pronounced (e.g., the latter portion of the 1998/1999 outburst), based on 
the work of Sobczak et al. (2000). For H~1743--322, we selected as many 
observations as necessary to make sure that the spectral evolution of 
H~1743--322 is adequately covered throughout each of its outbursts.

We followed our usual procedure to reduce and analyse the PCA and HEXTE data 
collected in the standard modes (e.g., Cui 2004, Xue \& Cui 
2005)\footnote{also see 
http://heasarc.gsfc.nasa.gov/docs/xte/recipes/cook\_book.html}. The data were 
reduced with {\em FTOOLS 5.2}. A PCA or HEXTE spectrum consists of separate 
spectra from individual detector units that were in operation. In deriving 
the PCA spectra, we only used data from the first xenon layer of each detector
unit (which is best calibrated). To estimate the PCA background, we used the 
background model for bright sources (pca\_bkgd\_cmbrightvle\_eMv20030330.mdl).

For each {\em RXTE} observation, we jointly fitted the PCA and HEXTE spectra
with a model that consists of a multi-color disc (`diskbb' in XSPEC) and a
power law with high-energy cutoff, taking into account the interstellar
absorption. We also introduced an additional multiplicative factor to account
for any uncalibrated difference in the overall throughput among
the individual detector units between the PCA and HEXTE. We limited the
PCA data to 3--30 keV and the HEXTE data to 15--200 keV and added 1\%
systematic uncertainty to the data. In some cases, we still needed an
additional Gaussian component to achieve statistically acceptable fits,
with its centroid at about 6--7 keV. The feature might be real (e.g., iron
fluorescence line) or an artefact due to inaccuracy in the calibration
around the xenon L edge (at $\sim$4.78 keV) and/or inadequacy of the model
(in the overlapping region of the two main components). Here, we are only 
interested in using the best-fitting model to unfold each observed (count) 
spectrum to derive the corresponding photon spectrum for further modelling.

\section{Jet Emission}

\subsection{XTE J1550--564}

A broadband SED of the western `blob' of XTE J1550--564 is shown in Fig.~1.
The X-ray data were taken from the {\em Chandra} observation conducted on
2002 March 11 (ObsID 3448; see Table~1). One important caveat is that the 
radio and X-ray measurements were not made simultaneously. In fact, they 
were made five days apart! This is relevant because the jet emission may be 
variable on a time-scale of days. With this caveat in mind, we proceeded with 
the assumption that the western `blob' did {\em not} vary appreciably at 
radio and X-ray frequencies in this case. We modelled the jet SED with a 
homogeneous synchrotron model (see Xue, Yuan \& Cui 2006 for details), 
as well as with an 
inverse Compton (IC) scattering model. For the latter, we estimated 
contributions from different sources of seed photons separately. We considered 
synchrotron self-Compton (SSC) scattering and IC scattering off the cosmic 
microwave background photons (IC/CMB). For low mass X-ray binaries, the photon
field of the accretion disc or of the companion star is probably much less 
important for the IC process in the jets.

\begin{figure}
\epsfig{figure=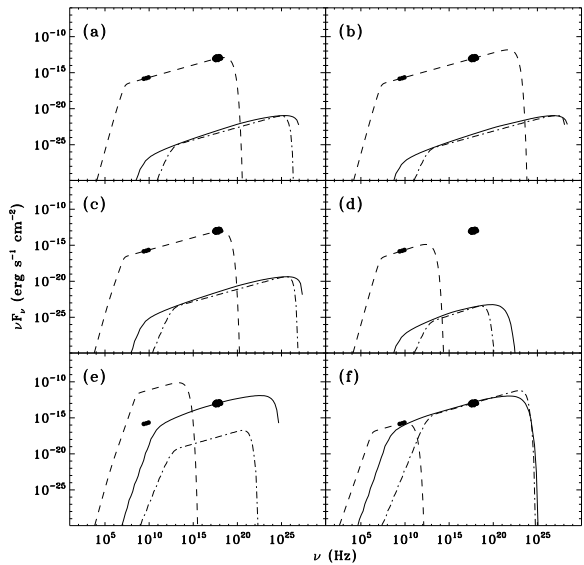, width=8.5cm, angle=0}
\caption{Spectral energy distribution of the western `blob' in the jets
of XTE J1550--564. The radio measurements were made with ATCA on MJD 52339
(Corbel et al. 2002). The X-ray measurements were made with {\em Chandra} on
MJD 52344. The dashed, solid, and dash-dot lines show contributions from
the synchrotron, SSC, and IC/CMB components, respectively.
See Table 3 for model parameters for each case.}
\end{figure}

From the {\em Chandra} X-ray image, we estimated the diameter of the western 
`blob' to be roughly 3\arcsec\ (though it is not exactly of spherical 
shape). Adopting the distance of $d=5.3$ kpc
(Orosz et al. 2002), we have the size of the emitting `blob'
$r \sim 1\times 10^{17}$ cm. Corbel et al. (2002) estimated the apparent
velocities of both eastern and western `blobs',
roughly 1.0 and 0.6 $c$, respectively. Assuming that the `blobs' were
ejected directly opposite to each other with the same velocity, we found
that the actual velocity was about 0.8 $c$ and the inclination angle of the
jet (with respect to the line of sight) about 72\arcdeg. The latter is
compatible with the optical measurements (Orosz et al. 2002). Therefore, the
Doppler factors of the eastern and western `blobs' are roughly 0.8 and 0.5,
respectively. It should be noted that the velocity of the `blobs' is
thought to be much larger early on (Hannikainen et al. 2001). For subsequent
modelling, we fixed the Doppler factor of the western `blob' at $\delta=0.5$.
Moreover, we fixed the minimum Lorentz factor of electron $\gamma_{\rm min}$
at 10, which is sufficiently low as not to affect the conclusions. The
remaining parameters in the model are: spectral index of electron $p$,
maximum Lorentz factor of electron $\gamma_{\rm max}$, energy density of
electron $E_{\rm tot}$, and magnetic field $B$.

Fig.~1 also shows representative fits to the data, 
with the corresponding model 
parameters summarized in Table~3. In general, the jet SED can be fitted well 
with synchrotron emission alone. We found that the electron spectral index
($p$) is well constrained, in the range of 2.28--2.32, but other parameters 
are much less so, $B$ in the range of 0.001--0.050 G, $\gamma_{\rm max} 
\gsim 1.26\times10^7$, and $E_{\rm tot}/m_e c^2$ in the range of 20--32000 
cm$^{-3}$, due partly to significant degeneracy 
in the fits. The best-fitting $p$ 
is 2.315, which is in good agreement with Corbel et al. (2002). In such cases 
(a--c), the IC (either SSC or IC/CMB) contribution to the measured fluxes 
is negligible. We also attempted to fit the radio data with synchrotron 
emission and the X-ray data with IC emission but failed to find any solutions.
Cases (d--f) in Fig.~1 show examples of the fits. While Case (f) appears to 
show a reasonable fit, the magnetic field required would be 
$8.0\times 10^{-9}$ G, which is unrealistically small for jets in microquasars.

\begin{table}
\caption{Model parameters for the western jet `blob' in XTE J1550--564}
\begin{tabular}{ccccc} \hline \hline
Case & $B$ & $E_{\rm tot}/m_ec^2$ & $\gamma_{\rm max}$ & $p$ \\
Number & (Gauss) & (cm$^{-3}$) & & \\ \hline
(a) & 0.016 & 25.1 & 1.6$\times 10^7$ & 2.31  \\
(b) & 0.032 & 79.5 & 5.0$\times 10^8$ & 2.32  \\
(c) & 0.002 & 1.2$\times 10^3$ & 3.0$\times 10^7$ & 2.32  \\
(d) & 0.032 & 0.06 & 1.0$\times 10^4$ & 2.20  \\
(e) & 0.003 & 4.0$\times 10^5$ & 1.0$\times 10^5$ & 2.30  \\
(f) & 8.0$\times 10^{-9}$ & 1.5$\times 10^{11}$ & 1.6$\times 10^6$ & 2.30 \\ \hline
\end{tabular}
\end{table}

\subsection{H 1743--322}

Similarly, we modelled a broadband SED of the eastern `blob' of H 1743--322,
as shown in Fig.~2. In this case, we estimated the diameter of the `blob' 
to be roughly 3.2\arcsec, based on the {\em Chandra} X-ray image, which
corresponds to a linear size of $r \sim 2\times 10^{17}$ cm, if we assume a 
distance of 8.0 kpc for the source. Corbel et al. (2005) found a solution to 
the intrinsic velocity of the `blobs' ($v = 0.79$ $c$) and the inclination 
angle of the jet ($\theta=73\arcdeg$), which would imply that the Doppler 
factor of the eastern `blob' is $\delta=0.8$. We fixed $\delta$, as well 
as the minimum Lorentz factor of the emitting electrons (at 
$\gamma_{\rm min}=10$) in the fits. Because the quality of the data is 
not as good for H 1743--322 as for XTE J1550--564, we also fixed the electron
spectral index ($p =2.20$; Corbel et al. 2005).

\begin{figure}
\epsfig{figure=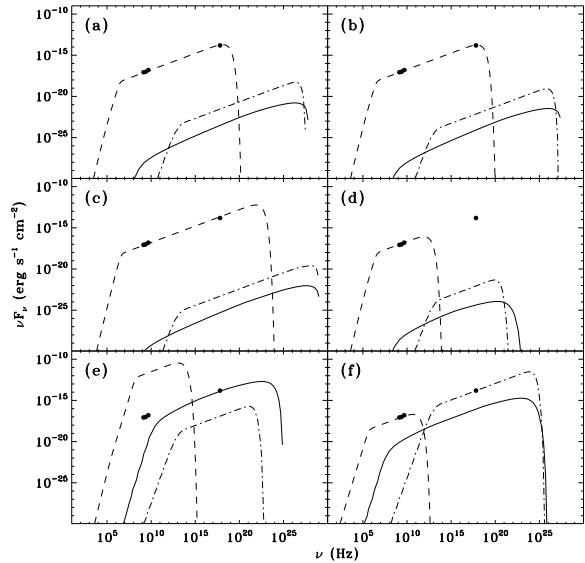, width=8.5cm, angle=0}
\caption{As in Fig. 1, but for the eastern `blob' in the jets of H 1743--322
on MJD 53047. See Table 4 for model parameters for each case.}
\end{figure}

Fig.~2 also shows representative fits to the data, with the corresponding model 
parameters summarized in Table~4. Like in the case of XTE J1550--564, 
the jet SED can, in general, be accounted for by synchrotron emission alone
(see Cases (a--c)). We also found that the X-ray emission of the jet could
not be explained by the IC emission either, were the observed radio fluxes 
attributed to synchrotron emission (see Cases (d--f)). 
An apparent good fit was achieved, as 
shown in Case (f), but the required magnetic field would be as small as 
$6.0\times 10^{-9}$ G, which is, again, unrealistic.

\begin{table}
\caption{Model parameters for the eastern jet `blob' in H 1743--322}
\begin{tabular}{ccccc} \hline \hline
Case & $B$ & $E_{\rm tot}/m_ec^2$ & $\gamma_{\rm max}$ & $p$ \\
Number & (Gauss) & (cm$^{-3}$) & & \\ \hline
(a) & 4.0$\times 10^{-4}$ & 344.0 & 5.0$\times 10^7$ & 2.20  \\
(b) & 0.001 & 55.0 & 2.5$\times 10^7$ & 2.20  \\
(c) & 0.005 & 71.0 & 1.0$\times 10^9$ & 2.20  \\
(d) & 0.001 & 0.25 & 2.5$\times 10^4$ & 2.20  \\
(e) & 0.001 & 9.9$\times 10^4$ & 1.0$\times 10^5$ & 2.20  \\
(f) & 6.0$\times 10^{-9}$ & 1.7$\times 10^{9}$ & 2.5$\times 10^6$ & 2.20 \\ \hline
\end{tabular} 
\end{table}

\section{Total Emission: Jet Contribution}

It is generally thought that radio emission from microquasars is due entirely
to the jets. A natural question is then how much the jets contribute to
emission at higher frequencies. The answer is invariably model-dependent
due to the lack of detailed understanding of the formation of the jets and
of the coupling between the jets and accretion flows. Here, we attempted
to address the issue by using some of the results obtained in the previous
section.

Fig.~3 shows the two best broadband SEDs of XTE~J1550--564 in our sample,
corresponding to two different spectral states (see discussion in \S~5). 
Assuming that the synchrotron-emitting electrons follow the same spectral 
energy distribution for all (resolved or unresolved) `blobs' in the jets 
(see evidence in Table~2), we fixed $p=2.32$ and 
$\gamma_{\rm max}=5.0\times10^8$, based on a solution for the 2002 western
`blob' (as shown in Fig.~1(b)). We also fixed the Doppler factor
at $\delta=0.5$ and the radius of the emitting region at $r=1\times10^{10}$ 
cm. A sufficiently small $r$ was adopted because the emission from the jets 
is likely dominated by optically-thick synchrotron emission from unresolved
components at the core (i.e., close to the black hole). We should note that 
the choice of the parameter values does not affect our general conclusions
(see below). The only remaining free parameters are $B$ and $E_{\rm tot}$.

\begin{figure}
\epsfig{figure=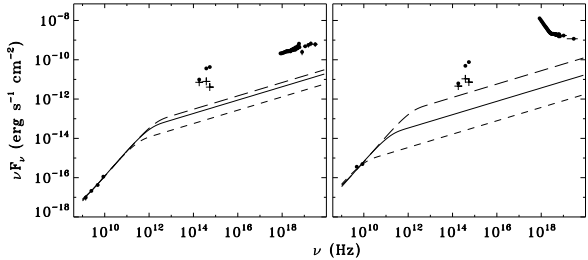, width=8.5cm, angle=0}
\caption{Spectral energy distribution of the total emission from
XTE J1550--564 in the low-hard state ($left$) and the transitional 
state ($right$). The filled circles show simultaneous/contemporaneous data
from radio (Corbel et al. 2001),
IR/optical (Jain et al. 2001), and X-ray measurements taken around
MJD~51696.47 ($left$) and MJD~51665.40 ($right$), respectively.
Note that the filled circles and pluses at the IR/optical
frequencies correspond to extinction corrections with $A_V=4.75$ (Orosz et al.
2002) and $A_V=2.2$ (S\'{a}nchez-Fern\'{a}ndez et al. 1999), respectively,
due to the uncertainty in the interstellar extinction for the source.
The short-dashed, solid, and
long-dashed lines show plausible solutions of the synchrotron model,
$left$: ($B=1778.7$ G, $E_{tot}/m_ec^2=1.1\times 10^{16}$ cm$^{-3}$),
($B=6311.3$ G, $E_{tot}/m_ec^2=4.5\times 10^{15}$ cm$^{-3}$),
and ($B=11223.3$ G, $E_{tot}/m_ec^2=2.8\times 10^{15}$ cm$^{-3}$);
$right$: ($B=56.2$ G, $E_{tot}/m_ec^2=1.0\times 10^{18}$ cm$^{-3}$),
($B=891.5$ G, $E_{tot}/m_ec^2=1.0\times 10^{17}$ cm$^{-3}$),
and ($B=7945.5$ G, $E_{tot}/m_ec^2=2.0\times 10^{16}$ cm$^{-3}$),
respectively, with $p=2.32$ and $\gamma_{\rm max}=5.0\times 10^8$ fixed
for all cases. Note that all plausible solutions fall between the
long-dashed and short-dashed lines.}
\end{figure}

We fitted the radio data with the synchrotron model. Some representative 
solutions are shown in Fig.~3. The spectral break at $\sim 10^{11-12}$ Hz 
marks the transition between the optically-thick and optically-thin 
synchrotron regimes. We note that no solutions were found under the
condition of equipartition between the electrons and the magnetic field,
which is in agreement with Wang, Dai \& Lu (2003). One thing that seems clear 
is that the synchrotron radiation from the jets contributes little to the 
observed optical and X-ray emission, at least in these two particular cases 
(when the source was relatively bright during the 2000 outburst). This 
conclusion does not appear to be sensitive to the adopted values of $p$ 
and $\gamma_{\rm max}$, because we also experimented with other values. 

Similarly for H 1743--322, we carried out similar modelling of the two best 
SEDs of the total emission in our sample, which are shown in Fig. 4,
also corresponding to two different spectral states.
In this case, we 
fixed the following parameters: $\delta=0.8$, $r=1\times10^{10}$ cm, 
and $p=2.20$. Representative solutions are also shown in Fig. 4.
Again, we found no solutions under the condition of equipartition between 
the electrons and the magnetic field. The results support the conclusion
that, while the jets can probably account for 100\% of the observed radio
emission, they contribute little to the observed emission at higher
frequencies, when the source is relatively bright.

\begin{figure}
\epsfig{figure=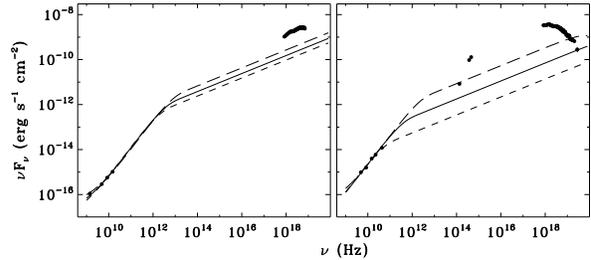, width=8.5cm, angle=0}
\caption{As in Fig. 3, but for H 1743--322 in the low-hard state ($left$)
and the transitional state ($right$). The data were taken around
MJD~52729.79 ($left$) and MJD~52733.78 ($right$), respectively,
$left$: radio data from Rupen et al. (2003), $right$: both radio and
IR/optical data from McClintock et al. (2007).
The short-dashed, solid, and
long-dashed lines show plausible solutions of the synchrotron model,
$left$: ($B=7945.5$ G, $\gamma_{\rm max}=1.4\times 10^7$, $E_{tot}/m_ec^2=7.9\times 10^{15}$ cm$^{-3}$), ($B=14129.3$ G, $\gamma_{\rm max}=1.4\times 10^8$, $E_{tot}/m_ec^2=3.2\times 10^{16}$ cm$^{-3}$), and
($B=31632.0$ G, $\gamma_{\rm max}=4.0\times 10^6$, $E_{tot}/m_ec^2=8.9\times 10^{14}$ cm$^{-3}$);
$right$: ($B=199.5$ G, $\gamma_{\rm max}=3.2\times 10^6$, $E_{tot}/m_ec^2=1.3\times 10^{17}$ cm$^{-3}$), ($B=1412.6$ G, $\gamma_{\rm max}=1.0\times 10^6$, $E_{tot}/m_ec^2=1.1\times 10^{16}$ cm$^{-3}$), and
($B=8913.2$ G, $\gamma_{\rm max}=1.0\times 10^5$, $E_{tot}/m_ec^2=4.5\times 10^{14}$ cm$^{-3}$), respectively, with $p=2.20$ fixed for all cases.}
\end{figure}

Where does the bulk of the X-ray emission come from then? We speculate that 
it originates mostly from accretion flows, as for XTE J1118+480 (Yuan et al. 
2005). Indeed, when we followed the usual empirical approach of modelling the 
X-ray spectrum, with a power law (plus a high-energy rollover for the 
low-hard state), which approximates an unsaturated Comptonized spectrum, 
and a multi-color disc component (when needed), we found that the model can
adequately fit the data. In principle, inverse Compton scattering could also
take place in the jets and the Comptonized photons could contribute to the
observed hard X-ray emission (Giannios 2005). However, we have shown in \S~3
that the IC emission from the jets is likely to play a negligible role, at 
least for XTE J1550--564 and H 1743--322.

On the other hand, as the source becomes fainter, the contribution of the
jets to the higher-frequency emission is thought to increase and perhaps 
eventually dominate (Yuan \& Cui 2005). To investigate whether there is
observational evidence to support it, we computed the ratio of the summed 
count rate, of all resolved jet `blobs' to the total count rate (of the 
core and `blobs') directly from the {\em Chandra} images of XTE J1550--564. 
The quantity should represent a {\em lower} limit on the fractional 
contribution of the jets to the total emission, since
there may be unresolved `blobs' along the jets or at the core. 
Fig.~5 show the results at 
various intensities of the source. It is clear that the jet emission can
indeed account for the bulk of, e.g., X-ray emission, when the source is
relatively faint. We were not able to repeat the analysis for H 1743--322 
due to the lack of statistics in this case.

\begin{figure}
\epsfig{figure=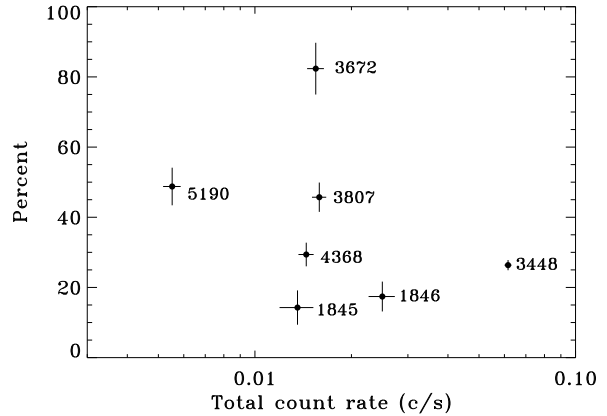, width=8.5cm, angle=0}
\caption{Contribution of the resolved jet `bolbs' in XTE J1550--564 to the
total observed X-ray emission, with observation ID shown for each point.
All the count rates were background-subtracted
and computed within the energy range of 0.3--8 keV. Note that each data point
only shows a {\em lower} limit on the overall contribution of the jets (see
text). }
\end{figure}

\section{Spectral Evolution}

The spectral evolution of microquasars is often described empirically in
terms of transitions between discrete spectral states (reviews by, e.g.,
Tanaka \& Lewin 1995; Liang 1998; Homan \& Belloni 2005; 
McClintock \& Remillard 2006). However,
different authors often define the states differently, which makes discussion 
and comparison of observational results difficult. In this work, we define
the low-hard state (LHS) as one in which the SED of the source peaks 
at $\sim 100$ keV and the high-soft state (HSS) as one in which the SED
peaks at $\sim 1$ keV. From the theoretical point of view, LHS thus defined
represents a physical configuration in which the X-ray spectrum is due 
entirely to
Comptonized emission (e.g., from hot accretion flows), while HSS represents 
one in which the X-ray spectrum is due entirely to optically-thick emission
(from cold accretion flows). As such, they would correspond to the most
diametrically opposed theoretical scenarios for microquasars. Strictly
speaking, therefore, one is likely to observe a source only in a quasi LHS
or HSS. During a transition between the two states, the source is
expected to show `intermediate' properties that one often observes in
the `intermediate state' or `very high state'. For transient microquasars,
it is also necessary to introduce the `quiescent state', which may or may
not be a simple extension of the LHS toward low fluxes. 

\subsection{XTE J1550--564}

Within the established context, we now discuss the observed spectral evolution
of XTE J1550--564 during outbursts in 1998--2000. We examined
the 2000 outburst first, because the outburst has a relatively simple profile
and it was well covered observationally both during its rising and decaying
phases. Fig.~6 shows representative SEDs of XTE J1550--564 during the
2000 outburst, along with an overview of the outburst based on the ASM data.
The SEDs clearly show spectral softening and hardening of the source as the
outburst proceeds. More specifically, it is apparent that the source underwent
an LHS-to-HSS transition (the shape of the SED goes from being high-frequency 
peaked to low-frequency peaked) somewhere between observations taken on 
MJD 51658.60 and on MJD 51662.17 (see Panel I) during the rising phase of the 
outburst and an HSS-to-LHS transition somewhere between observations taken on 
MJD 51678.45 and on MJD 51682.31 (see Panel II) during the decaying phase. 
This is at odds with the description of states during the same outburst by 
Corbel et al. (2001), who implied that the HSS was never reached in the 
outburst (see Fig.~1 of their paper). Even with such a large uncertainty in 
the timing of the transitions, we can still see evidence for spectral 
hysteresis associated with the rise and fall of the outburst, in the sense 
that the transitions seem to have occurred at different fluxes. The concept 
of spectral hysteresis is not new (Miyamoto et al. 1995; 
Nowak, Wilms \& Dove 2002; 
Maccarone \& Coppi 2003; Zdziarski et al. 2004). However, it can be confusing 
to discuss hysteresis without a clear definition of the states involved and a 
way to quantify the timing of the transitions. We will elaborate on this point
below.

\begin{figure}
\epsfig{figure=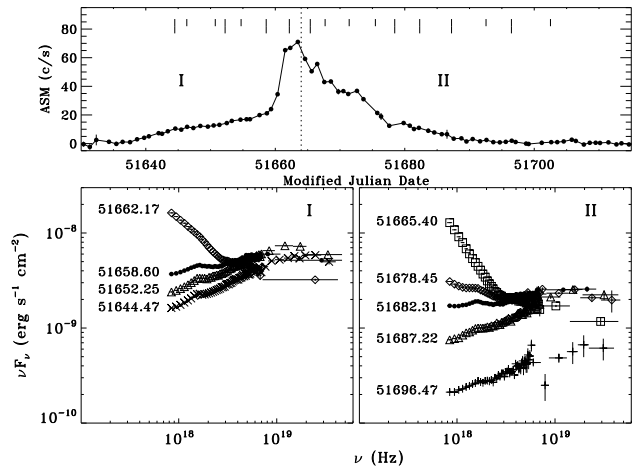, width=8.5cm, angle=0}
\caption{{\it Top}: Daily-averaged summed-band (1.5--12 keV) ASM/{\em RXTE}
light curve of XTE J1550--564 in 2000, with the rising and decaying phases
marked for comparison. Both short and long vertical lines indicate
the times of the pointed PCA/HEXTE observations we chose in this work,
with long vertical lines indicating representative SEDs shown in the
{\it Bottom} panels. {\it Bottom}: Representative SEDs for the rising and
decaying phases of the outburst, respectively. The start time (in MJD) of
each observation selected is indicated. }
\end{figure}

We repeated the analysis for the 1998/1999 outburst of XTE J1550--564, which 
is much stronger in its peak magnitude. Fig.~7 shows the representative 
SEDs, along with the ASM light curve. Spectral variability is also
apparent throughout the outburst. It is interesting to note that in this case 
the source never seems to have reached the HSS during the 1998 sub-outburst 
(corresponding to the first `hump' in the ASM light curve), judging from 
the shape of the SEDs. Instead, the SED of the source evolved from a typical 
LHS shape to an intermediate one that peaks at a significantly lower energy 
but without a dominant soft component (which defines the HSS SED). The soft 
component eventually emerged during the decaying phase of the sub-outburst 
(e.g., in the observation taken on MJD 51121.00, see Panel II). Spectral 
softening continued during the rising phase of the 1999 sub-outburst 
(corresponding to the second `hump' in the ASM light curve). The source 
finally reached the HSS at the peak of the sub-outburst (in the observation 
taken on MJD 51191.49, see Panel III). During the decaying phase of the
sub-outburst, the source went through an HSS-to-LHS transition (see Panel IV),
which is similar to the 2000 outburst but seems to have occurred at lower 
fluxes (comparing Panel II in Fig. 6 and Panel IV in Fig. 7). This supports 
the suggestion that spectral states cannot be uniquely determined by mass 
accretion rate (Homan et al. 2001).

\begin{figure}
\epsfig{figure=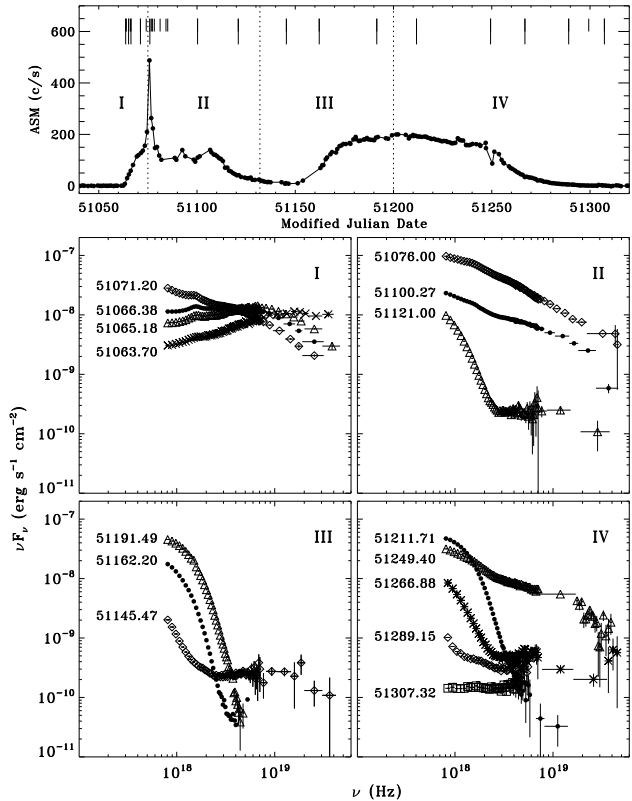, width=8.5cm, angle=0}
\caption{As in Fig.~6, but for XTE J1550--564 in the 1998/1999 outburst.}
\end{figure}

The SEDs shown have revealed that the HSS differs significantly from the 
`intermediate' period, which lasted for quite a long time during the 
1998/1999 outburst, although the `intermediate' SEDs seem to bear more 
resemblance to the HSS SED than the LHS SED. Maccarone
\& Coppi (2003) reported spectral hysteresis associated with the first
sub-outburst, based on the ASM light curve and hardness ratios. It is clear,
however, that that phenomenon is very different from the one that we are 
discussing here (i.e., spectral hysteresis associated with transitions 
between LHS and HSS), because the source was in constant transition and 
never in fact reached HSS during the sub-outburst. This
example illustrates the difficulty in comparing results in the literature
in the absence of a clear definition of the states.

\subsection{H~1743--322}

The representative SEDs of H~1743--322 are shown in Figs. 8--10 for its 2003, 
2004, and 2005 outbursts, respectively, along with the ASM light curves.
Interestingly, the 2003 outburst also has a two-hump profile, like the 
1998/1999 outburst of XTE J1550--564, while the 2004 and 2005 outbursts 
are more similar in shape to the 2000 outburst of XTE J1550--564. 

\begin{figure}
\epsfig{figure=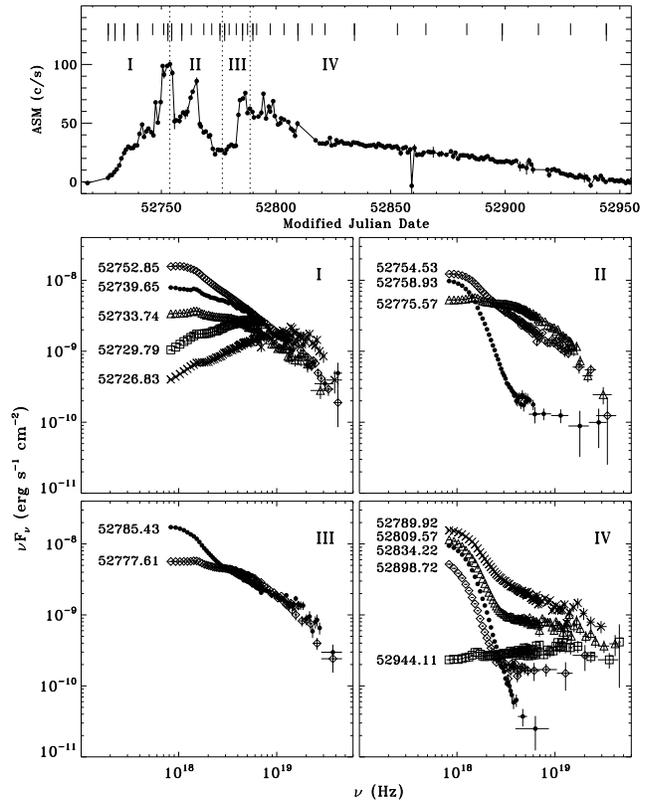, width=8.5cm, angle=0}
\caption{As in Fig.~6, but for H 1743--322 in the 2003 outburst.}
\end{figure}

\begin{figure}
\epsfig{figure=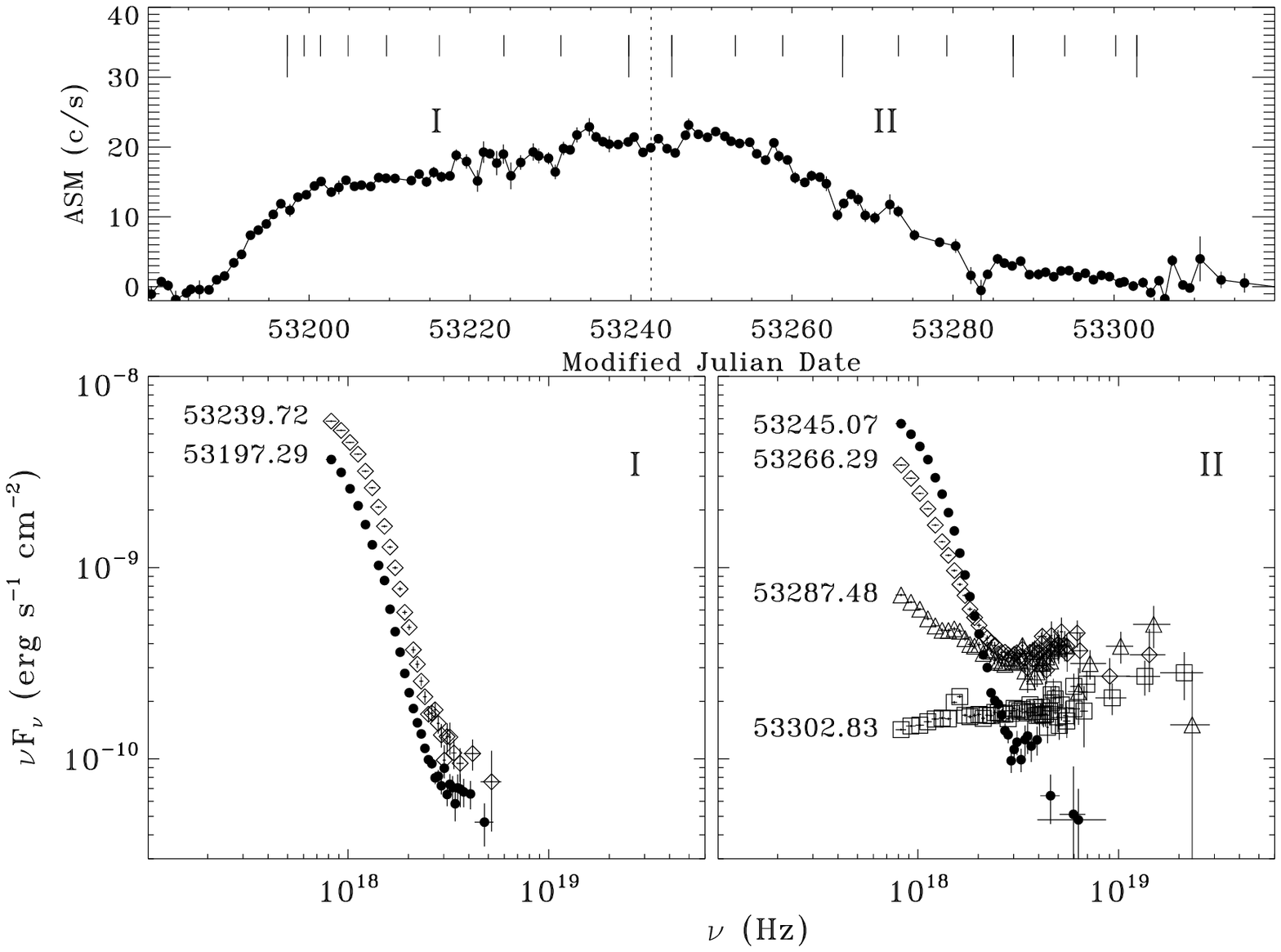, width=8.5cm, angle=0}
\caption{As in Fig.~8, but for H 1743--322 in the 2004 outburst.}
\end{figure}

\begin{figure}
\epsfig{figure=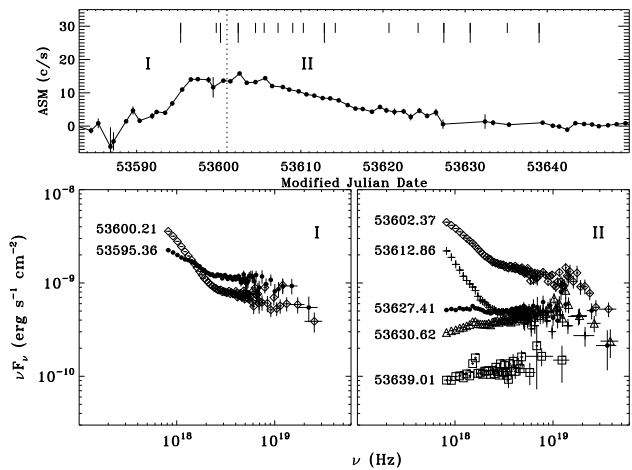, width=8.5cm, angle=0} 
\caption{As in Fig.~8, but for H 1743--322 in the 2005 outburst.}
\end{figure}

In the 2003 outburst (Fig.~8), during the rising phase of the first `hump' 
(Panel I), no HSS was reached, although the X-ray spectrum softened 
continuously. During the decaying phase of the first `hump' (Panel II) and 
the rising phase of the second `hump' (Panel III), the source exhibited 
complicated spectral behaviors: the soft (disc) component emerged, 
disappeared, and re-appeared. This is probably due to the sub-structures 
associated with both `humps'. The HSS was eventually reached during the 
decaying phase of the second `hump' (Panel IV), as indicated by the
dominance of the soft component. Subsequently, the source underwent an
HSS-to-LHS transition.

In the 2004 and 2005 outbursts (Figs. 9 and 10), the coverage of both rising
phases is not as good. No clear LHS was observed. However, HSS-to-LHS 
transitions can easily be inferred, during the decaying phases. Once again, 
we see that a specific state transition may occur at different fluxes 
(comparing Panel IV in Fig. 8, Panel II in Fig. 9, and Panel II in 
Fig. 10).

\section{X-ray/Radio Correlation}

Finally, we used the multiwavelength data to examine correlation between the
observed radio and X-ray variabilities. The correlation is of interest 
because it might be related to a coupling between the jets and accretion
flows in microquasars (Yuan \& Cui 2005). Such a coupling is a key ingredient
in many models of jet formation (e.g., Falcke \& Biermann 1995; Meier 2001).
The radio/X-ray correlation has been studied extensively (e.g., Corbel et al. 
2003; Gallo, Fender \& Pooley 2003; Fender, Belloni \& Gallo 2004; 
Xue \& Cui 2007). It was claimed that a `universal' radio/X-ray correlation 
exists for microquasars (Gallo, Fender \& Pooley 2003). However, the 
universality of the correlation has recently been ruled out by observations 
(Xue \& Cui 2007). It is nevertheless apparent that for a given microquasar, 
radio and X-ray variabilities may, to varying degrees, be correlated (which
can probably be extended to the infrared band; see Russell et al. 2007). To 
this end, we 
put together simultaneous/contemporaneous radio and X-ray measurements for 
XTE J1550--564. The results for H~1743--322 can be found in Xue \& Cui (2007).

Fig.~11 plots radio flux measurements ($F_R$), at 843, 4800, and 8640 MHz
respectively, against the 2--11 keV X-ray fluxes ($F_X$) of XTE J1550--564. 
Note that the
energy range for X-ray fluxes was chosen to facilitate direct comparison
with the published results (e.g., Xue \& Cui 2007). Although the
number of data points is quite limited, the data cover a dynamical range
of over two orders of magnitude both in radio and X-ray fluxes. At the first
glance, a general positive correlation is apparent. Following the literature,
we then fitted the data with a power law, in the form of
${\rm log_{10}}F_R=k{\rm log_{10}}F_X+b$, separately
for radio data at different frequencies. The results are also shown in
Fig.~11. The best-fitting logarithmic slope ($k$) 
was found to be $1.45$, $0.94$,
and $0.89$ for radio measurements at 843, 4800, and 8640 MHz, respectively.
However, none of the fits is formally acceptable; large scatters are
apparent.

\begin{figure}
\epsfig{figure=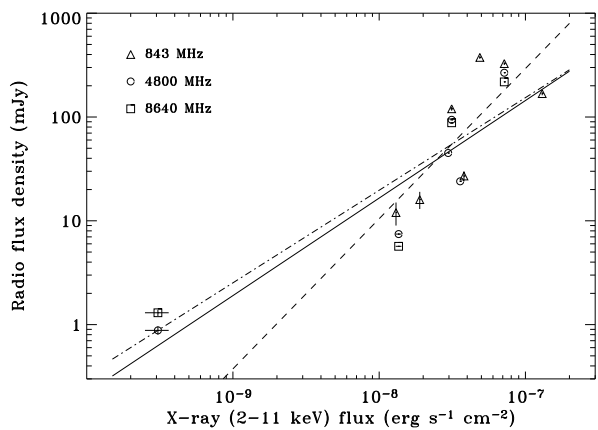, width=8.5cm, angle=0}
\caption{Radio/X-ray correlation in XTE J1550--564. The radio measurements
at 843, 4800, and 8640 MHz (taken from Wu et al. 2002, Hannikainen et al. 
2001, and Corbel et al. 2001) are shown in triangles, circles, and squares,
respectively. The dashed, solid, and dash-dot lines show separately the
best-fitting power laws to the data taken at 843, 4800, and 8640 MHz. Note
large scatters in the fits.}
\end{figure}

We note that the radio/X-ray correlation is often said to be applicable only 
to the LHS, in which the jets are thought to be compact and steady
(e.g., Corbel et al. 2001), but the 
radio/X-ray data used here are mostly for transitional periods between the LHS 
and HSS (when the jets are thought to be composed of discrete ejection 
events, e.g., Corbel et al. 2001), based on the definitions of the states 
adopted for this work. Yet, a 
general correlation seems to be apparent when {\it all} data points are 
included, at least for XTE~J1550--564, as shown in Fig.~11. Quantitative 
difference among the wavebands shows a possible spectral dependence of the correlation. 

\section{Summary} 

We have carried out a systematic study of the SED of XTE~J1550--564 and 
H 1743--322, and their spectral evolution during a number of major outbursts.
The main conclusions of the work are summarized as follows:

\begin{enumerate}
\item The results from physical modelling of the broadband SED of the resolved
components of the jets support a pure synchrotron origin of the observed 
emission, from radio to X-ray frequencies, in XTE J1550--564 and H 1743--322.
The effects of inverse Compton scattering in the jets were examined and found 
to be negligible. This is at variance with some of the published works (e.g., 
Markoff \& Nowak 2004; Giannios 2005; Markoff, Nowak \& Wilms 2005), which
argue for a dominant role of inverse Compton scattering in the production of
X-rays from microquasars based on modeling the unresolved (jets plus
core) emission of the sources.

\item We found that the synchrotron radiation from the jets can account for 
100\% of the observed radio emission but seems to contribute little to 
the observed X-ray emission from XTE J1550--564 and H 1743--322, when the 
sources are relatively bright. We think that the bulk of the X-ray emission 
comes from the accretion flows, which is always thought to be the case. 
We have found observational evidence to 
show that the jet contribution at X-ray energies increases as the sources 
become fainter, and might eventually dominate in or close to the quiescent 
state.

\item We found it straightforward to define the spectral states based 
on the shape of SEDs. The presence of a dominant soft (disc) component 
distinguishes the HSS from transitional states. In the context of the adopted 
definitions of the spectral states, we presented clear evidence for spectral 
hysteresis associated with the LHS-to-HSS and HSS-to-LHS transitions 
associated with the rising and decaying phases of an outburst. We also 
showed additional evidence to support a previous suggestion that the mass 
accretion rate alone cannot uniquely determine spectral states.

\item There is a general positive correlation between the X-ray and radio
fluxes in XTE J1550--564 even during the transitional states. We found 
evidence for a possible frequency dependence of the correlation.

\end{enumerate}

\section*{Acknowledgments}
We thank Feng Yuan for the use of his synchrotron utility and for helpful
discussions. Y.X. wishes to thank Keith Arnaud for helpful discussions
on part of the XSPEC modelling. 
W.C. also wishes to thank the colleagues in the Department of 
Astronomy at Peking University for their hospitality during his stay there. 
This work has made use of data obtained through the High Energy Astrophysics 
Science Archive Research Center Online Service, provided by the NASA/Goddard 
Space Flight Center. We gratefully acknowledge financial support from the 
Purdue Research Foundation, from the U.S. National Aeronautics and Space 
Administration through the ADP grant NNX07AH43G, from the National 
Natural Science Foundation of China through grants 10473001 and 10525313, 
and from the Ministry of Education (P.R.C.) through the RFDP Grant 
(No. 20050001026), the Key Grant
(No. 305001) and the NCET program (NCET-04-0022).


\begin{thebibliography}{99}
\bibitem[]{} Arnaud K. A., 1996, Astronomical Data Analysis Software and
Systems V, eds. Jacoby G. and Barnes J., ASP Conf. Series, vol. 101, p. 17
\bibitem[]{} Chaty S. et al., 2003, MNRAS, 346, 689
\bibitem[]{} Churazov E. et al., 1996, ApJ, 471, 673
\bibitem[]{} Corbel S. et al., 2001, ApJ, 554, 43 
\bibitem[]{} Corbel S. et al., 2002, Science, 298, 196
\bibitem[]{} Corbel S. et al., 2003, A\&A, 400, 1007
\bibitem[]{} Corbel S. et al., 2005, ApJ, 632, 504
\bibitem[]{} Corbel S. et al., 2006, ApJ, 636, 971
\bibitem[]{} Cui W., Zhang S. N., Chen W., Morgan E. H., 1999, ApJ, 512, L43
\bibitem[]{} Cui~W., 2004, ApJ, 605, 662
\bibitem[]{} Cui~W., 2006, Science, 309, 714
\bibitem[]{} Dickey J. M., Lockman F. J., 1990, ARA\&A, 28, 215
\bibitem[]{} Dove J. B.,  Wilms J., Begelman M. C., 1997, ApJ, 487, 747
\bibitem[]{} Esin A. A., McClintock J. E., Narayan R., 1997, 489, 865
\bibitem[]{} Falcke H., Biermann P. L., 1995, A\&A, 293, 665
\bibitem[]{} Fender R. P., 2006, in Compact Stellar X-ray Sources, eds. W. H. G. Lewin, M. van der Klis, (Cambridge: Cambridge Univ. Press), 381
\bibitem[]{} Fender R. P., Belloni T. M., Gallo E., 2004, MNRAS, 355, 1105
\bibitem[]{} Fuchs Y. et al., 2003, A\&A, 409, L35 
\bibitem[]{} Gallo E., Fender R. P., Pooley G. G., 2003, MNRAS, 344, 60
\bibitem[]{} Giannios D., 2005, A\&A, 437, 1007
\bibitem[]{} Grove J. E. et al., 1998, ApJ, 500, 899
\bibitem[]{} Hannikainen D. et al., 2001, Ap\&SS Suppl., 276, 45
%\bibitem[]{} Heinz S., Sunyaev R. A., 2003, MNRAS, 343, L59
\bibitem[]{} Hjellming R.~M., Han X., 1995, in X-ray Binaries, eds.
W. H. G. Lewin, J. Van Paradijs, E. P. J. van den Heuvel
(Cambridge: Cambridge Univ. Press), 308
\bibitem[]{} Homan J. et al., 2001, ApJS, 132, 377
\bibitem[]{} Homan J., Belloni T., 2005, Ap\&SS, 300, 107
\bibitem[]{} Hynes R. I. et al., 2002, MNRAS, 331, 169
\bibitem[]{} Jain R. K. et al., 2001, ApJ, 554, L181
\bibitem[]{} Kaaret P. et al., 2003, ApJ, 582, 945
%\bibitem[]{} Kaiser C. R., Sunyaev R., Spruit H. C., 2000, A\&A, 356, 975
\bibitem[]{} Liang E. P., 1998, Phys. Rep., 302, 67
\bibitem[]{} Maccarone T. J., Coppi P. S., 2003, MNRAS, 338, 189
\bibitem[]{} Markoff S., Falcke H., Fender R., 2001, A\&A, 372, L25
\bibitem[]{} Markoff S., Nowak M. A., 2004, ApJ, 609, 972
\bibitem[]{} Markoff S., Nowak M. A., Wilms J., 2005, ApJ, 635, 1203
\bibitem[]{} McClintock J.~E., Remillard R., 2006, in Compact Stellar X-ray
Sources, eds. W. H. G. Lewin, M. van der Klis, (Cambridge: Cambridge Univ. Press), 157
\bibitem[]{} McClintock J.~E. et al., 2007 (astro-ph/0705.1034)
\bibitem[]{} Meier D. L., 2001, ApJ, 548, L9
\bibitem[]{} Migliari S. et al., 2007, ApJ, in press (astro-ph/0707.4500)
\bibitem[]{} Mirabel I., 2004, PThPS, 155, 71
\bibitem[]{} Miyamoto S. et al., 1995, ApJ, 442, L13
\bibitem[]{} Morrison R., McCammon D., 1983, ApJ, 270, 119
\bibitem[]{} Narayan R., Mahadevan R., Quataert E., 1998, in The Theory of
Black Hole Accretion Discs, eds. M. A. Abramowicz, G. Bjornsson,
and J. E. Pringle (astro-ph/9803141)
\bibitem[]{} Nowak M. A, Wilms J., Dove J. B., 2002, MNRAS, 332, 856
\bibitem[]{} Orosz J. A. et al., 2002, ApJ, 568, 845
\bibitem[]{} Poutanen~J., Fabian~A.~C., 1999, A\&A, 306, L31
%\bibitem[]{} Remillard R. A., McClintock J. E., 2006, ARA\&A, 44, 49
\bibitem[]{} Rupen M. P., Mioduszewski A. J., Dhawan V., 2003, IAU Circ. 8105, 3
\bibitem[]{} Russell D. M. et al., 2007, MNRAS, 379, 1401
\bibitem[]{} S\'{a}nchez-Fern\'{a}ndez C. et al., 1999, A\&A, 348, L9
\bibitem[]{} Sobczak G. J. et al., 2000, ApJ, 544, 993
\bibitem[]{} Tanaka Y., Lewin W. H. G., 1995, in X-ray Binaries, eds.
W. H. G. Lewin, J. Van Paradijs, E. P. J. van den Heuvel (Cambridge: Cambridge Univ. Press), 126
\bibitem[]{} Titarchuk~L., Shrader~C.~R., 2002, ApJ, 567, 1057
\bibitem[]{} Tomsick J. A. et al., 2003, ApJ, 582, 933
\bibitem[]{} Ueda Y. et al., 2002, ApJ, 571, 918
\bibitem[]{} Wang X. Y., Dai Z. G., Lu T., 2003, ApJ, 592, 347
\bibitem[]{} Wu K. et al., 2002, ApJ, 565, 1161
\bibitem[]{} Xue Y. Q., Cui W., 2005, ApJ, 622, 160
\bibitem[]{} Xue Y. Q., Cui W., 2007, A\&A, 466, 1053
\bibitem[]{} Xue Y. Q., Yuan F., Cui W., 2006, ApJ, 647, 194
\bibitem[]{} Yuan F., Cui W., 2005, ApJ, 629, 408
\bibitem[]{} Yuan F., Cui W., Narayan R., 2005, ApJ, 620, 905
\bibitem[]{} Zdziarski~A.~A. et al., 2001, ApJ, 554, L45
\bibitem[]{} Zdziarski A.~A. et al., 2004, MNRAS, 351, 791
\end{thebibliography}
\end{document}